\documentclass[aps,prl,showpacs,floatfix,superscriptaddress,twocolumn]{revtex4-1}

\usepackage{bm}
\usepackage{amsmath,amsfonts,amssymb,graphicx}
\usepackage{epstopdf}
\usepackage{color}
\newcommand{\SiO}{SiO$_2$}
\newcommand*{\TLV}{Raymond and Beverly Sackler School of Physics and Astronomy, Tel-Aviv University, Tel Aviv 69978, Israel}
\newcommand*{\WIS}{Department of Condensed Matter Physics, Weizmann Institute of Science, Rehovot, 76100, Israel}
\newcommand{\beginsupplement}{%
        \setcounter{table}{0}
        \renewcommand{\thetable}{S\arabic{table}}%
        \setcounter{figure}{0}
        \renewcommand{\thefigure}{S\arabic{figure}}%
     }

\begin{document}

\title{Suppression of Coulomb blockade peaks by electronic correlations in InAs nanowires}

\author{R.~Hevroni}
\affiliation{\TLV}
\author{V.~Shelukhin}
\affiliation{\TLV}
\author{M.~Karpovski}
\affiliation{\TLV}	
\author{M.~Goldstein}
\affiliation{\TLV}	
\author{E.~Sela}
\affiliation{\TLV}	
\author{Hadas~Shtrikman}
\affiliation{\WIS}	
\author{A.~Palevski}
\affiliation{\TLV}


\date{\today}

\begin{abstract}
We performed electronic transport measurements on 1D InAs quantum wires.
In sufficiently disordered wires, transport is dominated by Coulomb blockade, and the conductance can be well described by tunneling through a quantum dot embedded between two one-dimensional Luttinger liquid wires.
In contrast to previous experiments in other material systems, in our system the conductance difference between peak to valley \emph{decreases} with decreasing temperature for several consecutive peaks.
This phenomenon is theoretically expected to occur only for strongly interacting systems with small Luttinger interaction parameter $g<1/2$; we find for our InAs wires a value of $g\approx{0.4}$. Possible mechanisms leading to these strong correlations are discussed.
\end{abstract}


\maketitle

Extensive theoretical \cite{paper1,paper2}  and experimental \cite{paper3,paper4,paper5} research has been devoted to one-dimensional (1D) InAs and InSb semiconductor nanowires (NWs) over the past few years. Such NWs, when put in proximity to a superconductor, are expected to host Majorana topological states due to strong spin-orbit coupling. The lack of a Schottky barrier with the metallic contact and the relatively large Land\'{e} factor make the experimental conditions for observing these exotic topological states quite feasible.  Indeed, a zero bias peak as a signature of Majorana states was reported in InSb \cite{paper4,paper5}  and InAs \cite{paper3} NWs in proximity to superconducting Nb and Al films, respectively.

One of the main features in such 1D wires is the existence of ballistic helical states which are theoretically predicted~\cite{paper6} to exhibit nonmonotonic (up and down) conductance steps of size $G_{0}=e^{2}/h$  as the electron density is varied by the gate voltage.
Although many attempts have been made to measure these conductance steps, they have not been observed yet in either InAs or InSb NWs. This indicates that disorder plays an essential role, preventing the motion of the electrons between the contacts from being ballistic. It is well known that in 1D electron-electron interactions, described by the Luttinger-liquid (LL) model, amplify the role of disorder significantly, causing the conductance to vanish at zero temperature even for very weak disorder \cite{paper8}. 
Experimentally, however, the effects of the interactions in NWs with strong spin-orbital scattering have not yet been reported.

In this letter we report on experimental studies of electronic transport in disordered InAs NW at low temperatures over a wide range of electron densities. At very low densities we find the transport to be governed by Coulomb blockade, and at relatively high electron density by sequential tunneling through a series of barriers present in the disordered NW. We demonstrate that in both regimes the conductance is strongly affected by electron-electron interactions. The analysis of the temperature dependence of the conductance and of the line shape of the resonant tunneling in the Coulomb blockade regime within the framework of the existing theories \cite{paper8} allows us to deduce the corresponding LL parameter $g$. We show that in our NWs the effective LL parameter reaches a value less than $1/2$, leading to a \emph{decrease} in the Coulomb blockade peak to valley difference as the temperature is reduced.
To the best of our knowledge, this phenomenon, predicted by the LL model, has never been experimentally observed before: While there were a number of experimental papers \cite{Meirav,Moser} in which the Coulomb peaks decreased with decreasing temperature, this behavior was sporadic, namely did not occur for consecutive peaks. Thus these previous results do not follow the predictions of the LL theory, but are rather consistent with stochastic Coulomb blockade \cite{Glazman}, while the opposite is true for our results, as we discussed below.

InAs NWs  approximately 2~$\mu${m} long and 50~nm in diameter were grown by Au-assisted vapor-liquid-solid molecular beam epitaxy on a 2{'}{'} \SiO/Si substrate. A $\sim$1~nm gold layer was evaporated in situ in a chamber attached to the MBE growth chamber after degasing of the substrate at 600~$^{\circ}$C. The substrate was heated to 550~$^{\circ}$C after being transferred to the growth chamber to form gold droplets, then cooled down to the growth temperature of 450~$^{\circ}$C.
Indium and As$_{4}$ were evaporated at a V/III ratio of 100. The NWs were studied by SEM and TEM and were found to have a uniform morphology with no tapering and a pure wurtzite structure with a negligible number of stacking faults \cite{paper9}. The NWs were deposited randomly from an ethanol suspension onto 300 nm thick \SiO\ thermally grown on a p$^{+}$-Si substrate, to be used as a back gate. The NW were then mapped with respect to alignment marks using an optical microscopy. Ti/Al (5~nm/100~nm) contact leads, 650 nm separated, were deposited on the NWs, using electron beam lithography and electron beam evaporation, see Fig.~\ref{fig1}. A short dip in an ammonium polysulfide solution \cite{Ammoniumsolfide} was used for removing the oxide from the InAs NWs surface prior to contacts deposition.

InAs NWs are highly sensitive to surface impurities and other imperfections (such
as surface steps and dangling bonds) since its conductance electrons are near the surface. Hence, impurities resulting from sample fabrication and the external environment, as well as the substrate on which the sample is placed, induce disorder potential barriers.

\begin{figure}{}
\centering
\includegraphics[width=0.4\textwidth]{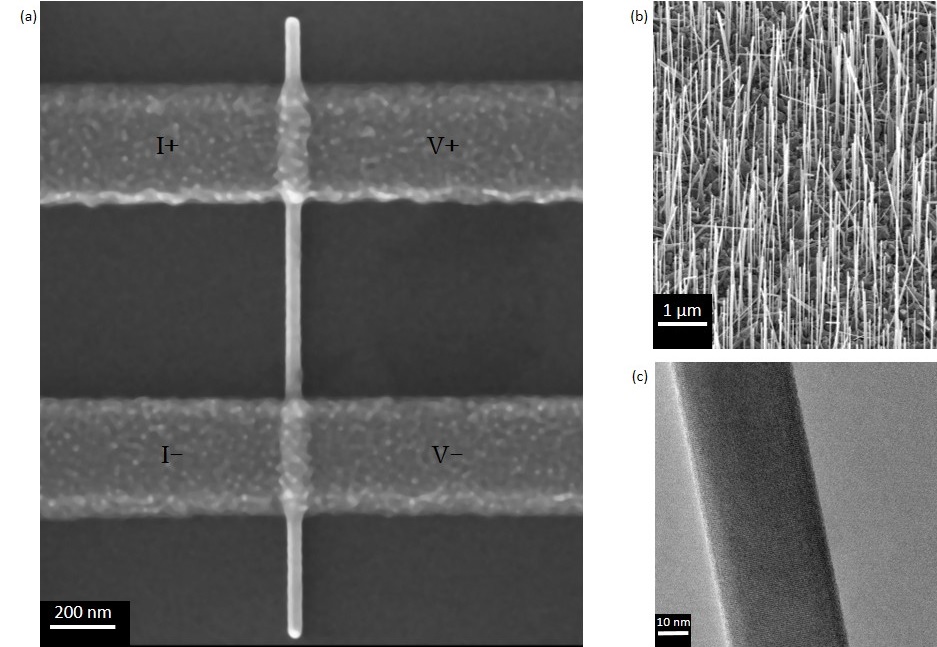}
\caption{(a) SEM image of a typical sample. The conductance was measured by
the four-terminal method, in which the current was passed between two
probes (denoted as I$+$ and I$-$), while the voltage was measured between
two other probes (V$+$ and V$-$).
(b) SEM image of InAs NWs as grown on \SiO/Si substrate.
(c) TEM image of an InAs NW. }\label{fig1}
\end{figure}

Conductance was measured by a four-terminal method using a low-noise analog lock-in amplifier (EG\&GPR-124A). The current was passed between two probes (I$+$ and I$-$ in Fig.~\ref{fig1}), while the voltage was measured by two different probes (V$+$ and V$-$ in Fig.~\ref{fig1}). It should be noted that I and V are connected to the NW at the same point, so that the contact resistance is always included in the conductance measured. The measurements were done in a $^{4}$He cryogenic system in a temperature range of 1.7~K--4.2~K.

Fig.~\ref{fig2} shows the measured conductance $G$ as a function of gate voltage, $V_{g}$, of an InAs NW over a wide range of gate bias. The as-grown NWs are conducting and the gate voltage bias required to pinch off the conductance is $V_{g}=-0.35$~V. At low values of the gate voltage (left panel) a series of distinct conductance peaks is clearly observed, with a typical spacing of $\delta{V_{g}}\sim$25~mV. At high gate voltage values (exceeding 0.5~V, right panel), the variation of the conductance is much smoother, with an overall tendency to increase with increasing gate voltage. At lower temperatures the conductance peaks become sharper, but the value of the conductance at the peaks is reduced.

\begin{figure}{}
\centering
\includegraphics[width=0.5\textwidth]{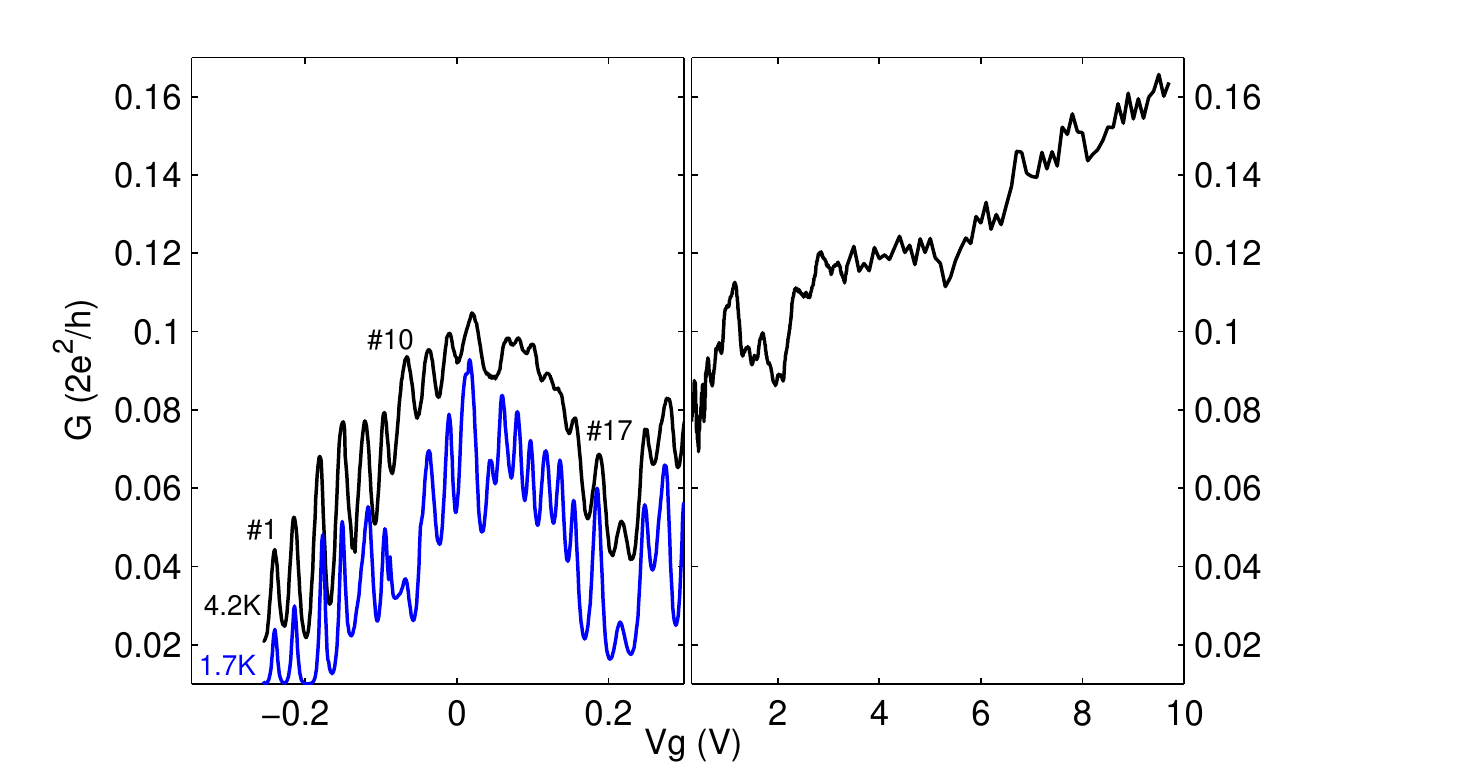}
\caption{(Color online) Conductance of an InAs NW (diameter: 50~nm; length: 650~nm) as function of the gate voltage, at $4.2$~K (black curve) and $1.7$~K (blue curve). The first conductance peak in marked by \#1, and the tenth by \#10.}\label{fig2}
\end{figure}

The behavior at low values of the gate voltage (left panel of Fig.~\ref{fig2}) indicates the occurrence of a Coulomb blockade, similar to quantum dots. The peak spacing in our device is shown in Fig.~\ref{fig3}.
It is well known that the conductance peaks in quantum dots are equally spaced if the energy level spacing between single electron states in the dot, $\Delta$, is negligible relative to the charging energy, and could become irregularly-spaced in the opposite limit \cite{QDreview}. Since here the distance between peaks varies by over 50\%, the latter case is realized in our sample. 
In this regime every peak corresponding to an odd number of electrons in the dot should be separated from the previous one by a roughly constant charging energy $E_{c}$, whereas the next peak should be separated by $E_{c}+\Delta$, and thus vary from level to level. Indeed, we see that every second peak of the first 10 peaks has a gate voltage spacing of 25~mV (with the exception of the distance between the 5th and 6th peaks which is slightly lower ($\approx$21~mV).
This value of $\delta{V_{g}} = 25$~mV should thus correspond to the charging energy, and is related to gate capacitance $C_{g}$ via $\delta{V_{g}}=e/C_{g}$, yielding a value of $C_{g}=6.4\cdot10^{-18}$ F.

\begin{figure}{}
\centering
\includegraphics[width=0.4\textwidth]{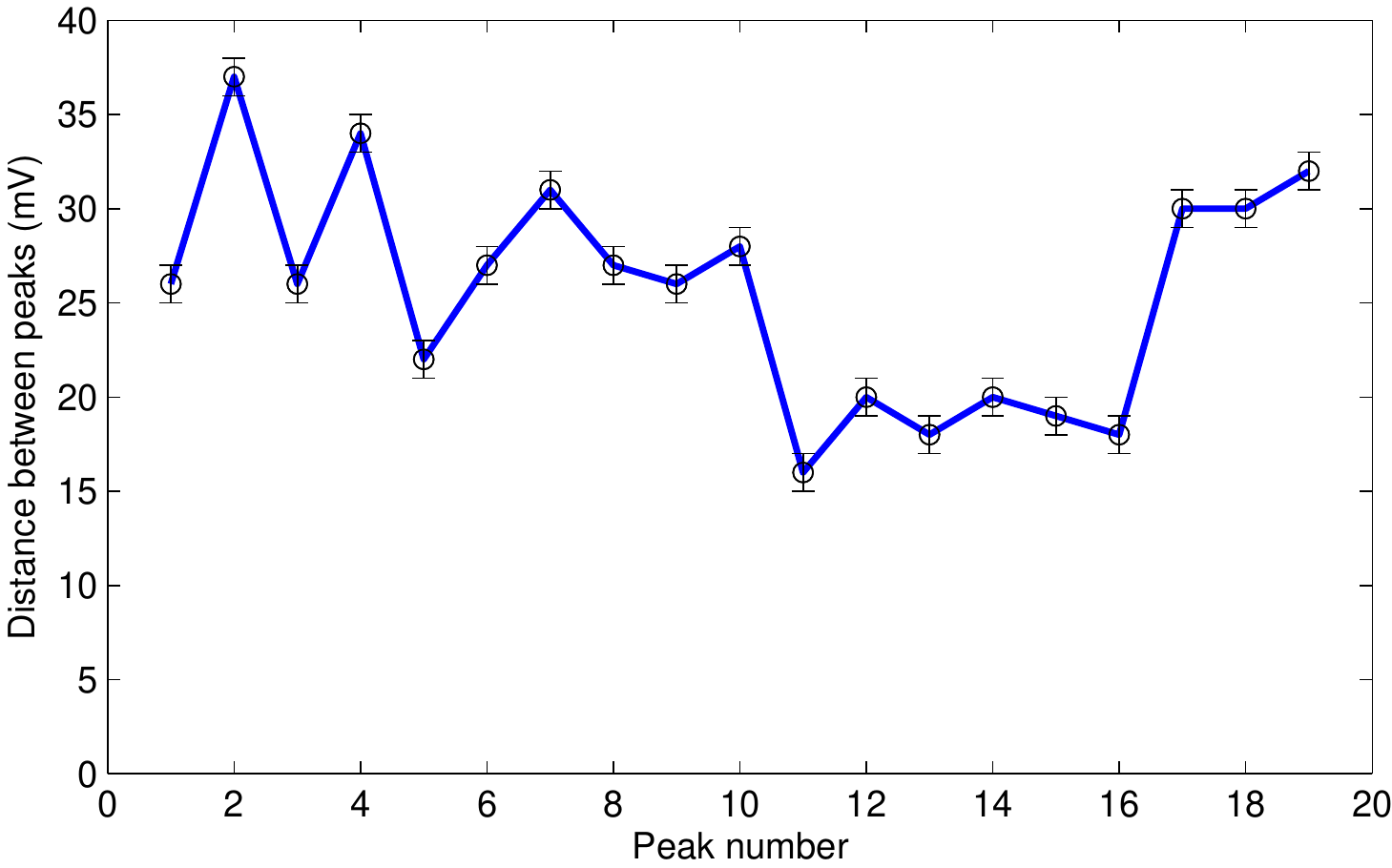}
\caption{The peak distance in mV. Its value at peak number n indicates the distance between peak \#n and peak \#n+1 taken at $T=1.7$~K.}\label{fig3}
\end{figure}

Since the geometry of the sample and its dimensions are known, we can estimate the size $L_{QD}$ of the quantum dot from the expressions for the capacitance of a cylinder in the vicinity of a conducting plate,
$C_g=2\pi\varepsilon{L_{QD}}/\ln(4d/D_\text{wire})$,
where $\varepsilon$ is the dielectric constant, $D_\text{wire}$ is the wire diameter, and $d$ is the \SiO~thickness.
Substituting the values of the sample dimensions, the average dielectric constant of $^{4}$He and \SiO~($\varepsilon \approx 2.5\varepsilon_{0}$), and the estimated value of the capacitance, gives $L_{QD} \approx 200$~nm.
We see that $L_{QD}$ is smaller than the NW length ($L= 2$~$\mu$m) by an order of magnitude, and smaller by more than a factor of 3 compared to the separation between the voltage leads ($L_{vp} = 650$~nm). Thus it is legitimate to assume that the QD is formed as a puddle of 1D electrons separated by two barriers on both its sides, somewhere in the segment of the NW between the voltage leads.
In addition, we see that the segments of the NW to the right and left of the dot are long enough so that their single particle level spacings and charging energies are well below the temperatures reached in our experiment, allowing us to describe them as infinite 1D leads.

In such a case it is reasonable to carry out our data analysis in the framework of the theory of tunneling between two 1D NWs through a quantum dot.
Resonant and sequential tunneling were well studied theoretically and experimentally in the past for both interacting and non-interacting 1D electrons, see \emph{e.g.}, Ref.~\cite{paper8} and references therein.
In our system the peaks widths are found to scale linearly with the temperature $T$.
We thus try to explain our results using Furusaki's expression \cite{paper10} for the conductance due to sequential electron tunneling in a QD connected to LL leads. The line shape of a single conductance peak as a function of the energy $E$ (distance from the peak), is then:
\begin{equation}
  G(E,T)=A G_{0}\frac{\gamma(T)}{T \cosh \left( \frac{E}{2k_{B}T} \right)} \left| \Gamma\left(\frac{1}{2g}+i\frac{E}{2\pi{k_{B}}T} \right)  \right |^{2},
\label{eq1}
\end{equation}
where $A$ is a constant related to the asymmetry and height of the barriers defining the dot, the factor $\gamma(T) = T^{1/g-1}$ accounts for the renormalization of the tunneling rates by the LL effects, and $\Gamma(z)$ is the gamma function. Note that the temperature variation of $G$ at the peak is then
\begin{equation}
G_\text{max}(T) \propto T^{1/g-2}. \label{eqn:gmax}
\end{equation}
In all the above expressions $g$ is the effective LL interaction parameter; $g=1$ for a non-interacting NW and decreases ($g<1$) with increasing repulsive interactions. It is a combination of the charge and spin interaction parameters, as we discuss below.

The experimental data in Fig.~\ref{fig2} shows that both the height and the width of the conductance peaks decrease as temperature is reduced. Thus the interaction parameter $g$ should be smaller than $1/2$. Our experimental data (Fig.~\ref{fig5}) shows that indeed the temperature dependence of the height of each peak can be well described by the power law, Eq.~(\ref{eqn:gmax}), from which we can deduce the value of $g$ for each peak. For the first two peaks we find $g = 0.38\pm0.03$.

\begin{figure}{}
\centering
\includegraphics[width=0.4\textwidth,height=0.25\textheight]{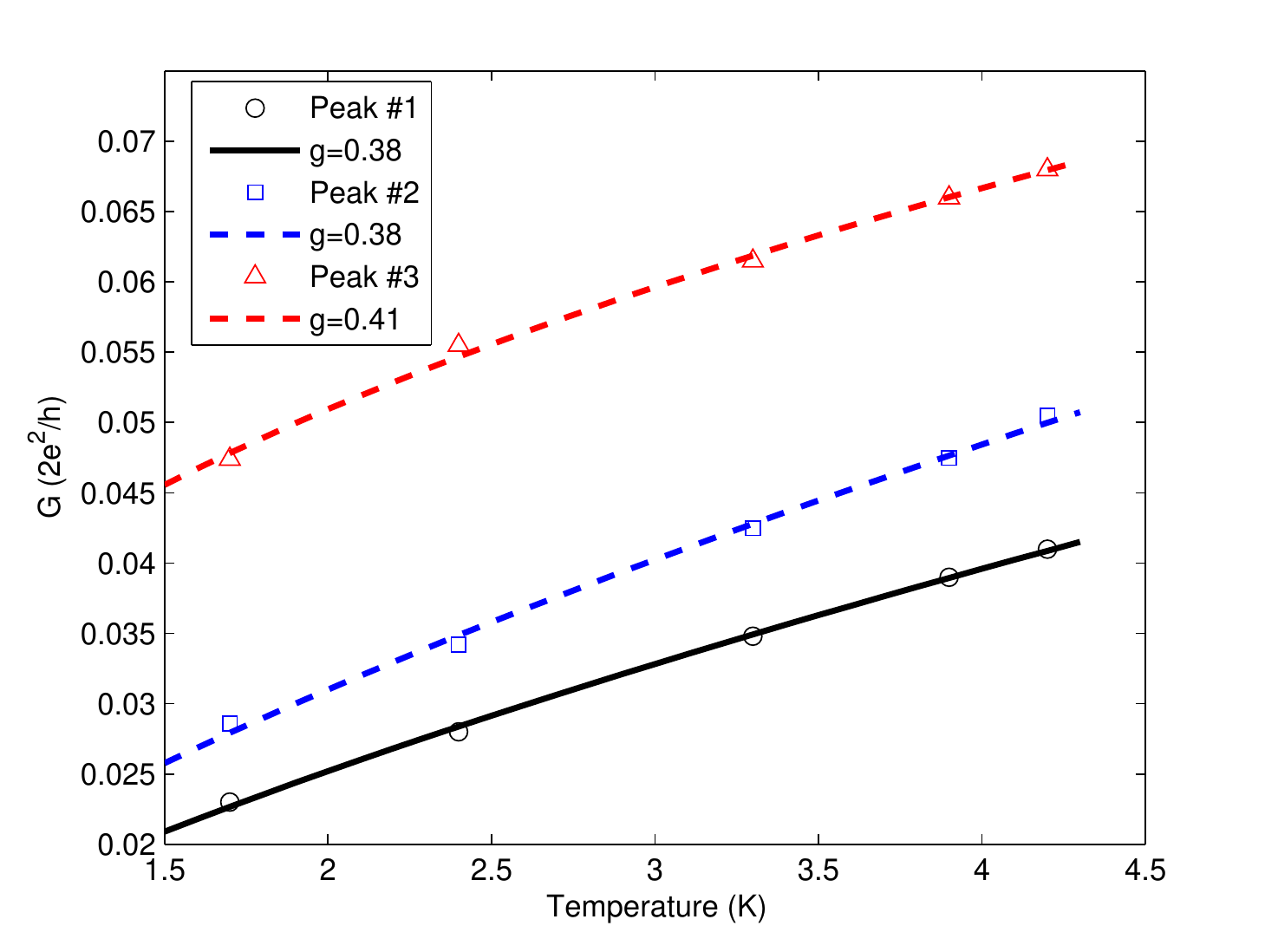}
\caption{(Color online) The experimental conductance peak heights as a function of temperature and fits to Eq.~(\ref{eqn:gmax}), for the first three peaks of Fig.~\ref{fig2}.}\label{fig5}
\end{figure}

\begin{figure}{}
\centering
\vspace{1pt}
\includegraphics[width=0.4\textwidth]{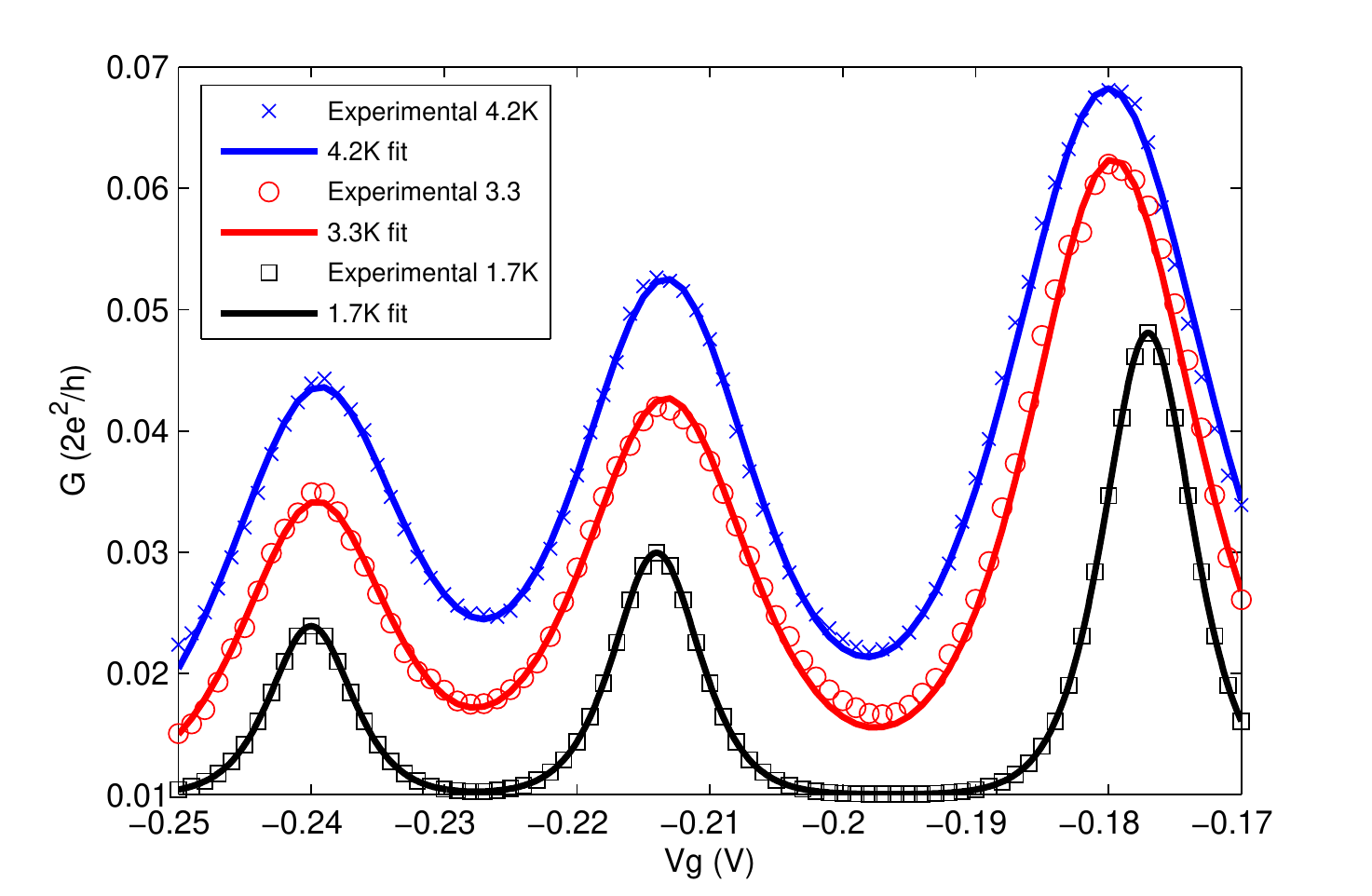}
\caption{(Color online) Fit of the first three peaks in Fig.~\ref{fig2} to Eq.~(\ref{eq1}) at 4.2~K, 3.3~K, and 1.7~K.}\label{fig4}
\end{figure}

In order to verify that the line-shape of the Coulomb blockade peaks as a function of the gate voltage can be described by expression~(\ref{eq1}), we fit the gate voltage dependence of our data to a sum of terms (one for each peak) of the form of Eq.~(\ref{eq1}),
with $E=\alpha(V_g-V_0)$, where $V_{0}$ is the gate voltage value at the peak and $\alpha$ is the ratio between the gate capacitance and the total capacitance of the dot. $\alpha$ and the amplitude $A$ are used as fitting parameters, and for $g$ we plug in the value extracted from the data analysis presented in Fig.~\ref{fig5}.
The result for the first 3 peaks is shown in Fig.~\ref{fig4}.
We find that, as expected, $\alpha = 0.1\pm0.005$ does not vary between the peaks and/or as function of temperature indicating that Eq.~(\ref{eq1}) indeed gives a consistent description of our dataset.
We have performed a similar fit procedure for other peaks, and in Figs.~S1--S2 of the supplementary material \cite{SMArchive} we show the results for peaks \#16--\#18. Since the Fermi energy is larger, the value of $g$ is expected to be higher for these peaks, and we indeed find $g\approx 0.5$.

We now turn to the analysis of the data at high gate voltages (right panel of Fig.~\ref{fig2}), where the Coulomb blockade oscillations are no longer observed. The conductance at this range of gate voltages also exhibits a power-law decrease as the temperature is decreased. Fig.~\ref{fig6} shows the experimental data for the temperature variation of the conductance at $V_{g}=3.5V$, together with a power-law fit. 
One can argue that in this regime the electronic conductance can be described by a model of a LL with strong disorder, in which the conductance is given by \cite{paper8}:
\begin{equation}
G(T) \propto T^{1/g-1}.
\label{eq3}
\end{equation}
From the fit we find that at high gate voltages $g \approx 0.74$, higher than the values extracted within Coulomb blockade regime for a much lower gate voltage. It is indeed expected that the interaction constant should increase as the Fermi energy is increased, since $g$ depends on the ratio between the Coulomb energy $U$
and the Fermi energy $E_{F}$ in the NW.
Moreover, we might have more than one conducting channel in the NW at such high gate voltage, which could also lead to the increase in the value of $g$.

\begin{figure}{}
\centering
\includegraphics[width=0.4\textwidth,height=0.25\textheight]{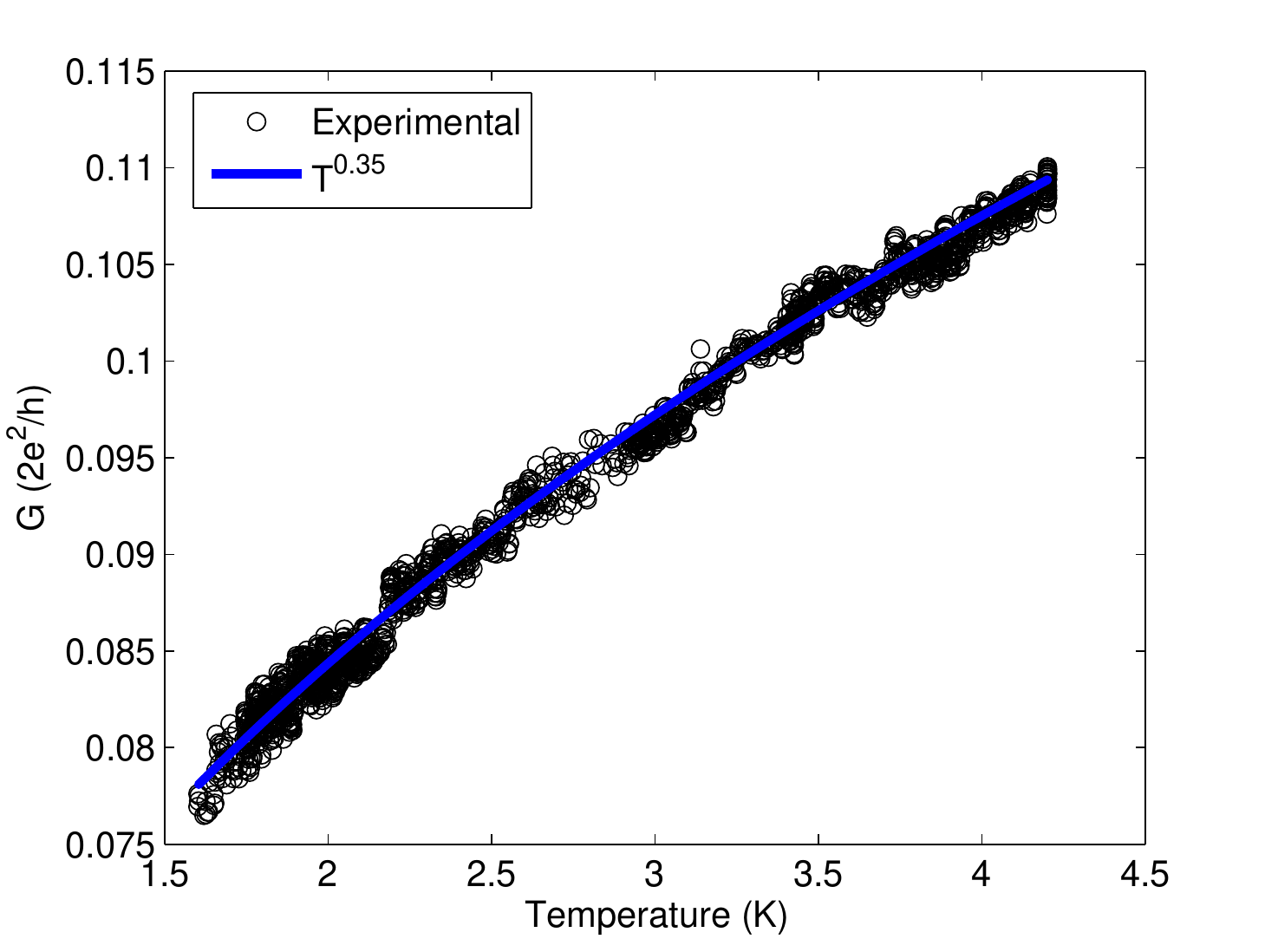}
\caption{(Color online) Fit to Eq. \ref{eq3} (solid blue curve) to the conductance vs. Temperature at $V_{g}=3.5V$}\label{fig6}
\end{figure}

Another interesting feature of the conductance variation at low gate voltage is the non-monotonic variation of the average conductance values: As one can see in Fig.~\ref{fig2}, while the average conductance  is generally  increasing along almost the entire range of the gate bias, a local minimum around  $V_{g}\approx0.2V$ is clearly observed. The monotonic
increase of the average conductance is expected, since the transmission of the potential barriers should increase with energy, and, moreover, electrons start to populate additional bands at higher gate bias. The well-pronounced minimum of the average conductance which is surprising.
One could speculate that it might be related to the opening of an energy gap in the higher bands due to disorder and strong spin-orbit coupling, as was recently predicted theoretically~\cite{Meng}. However, we cannot verify this scenario quantitatively.

As we pointed out earlier, the reduction of the conductance peaks at low temperatures has been observed in previous experiments \cite{Meirav,Moser}, but their results are markedly different than ours.
In previous experiments, the decrease was sporadic, occurring only for non-consecutive peaks, and thus cannot be accounted for by the LL picture, but rather indicates a stochastic Coulomb blockade \cite{Glazman}. In contrast, in our system the peak reduction occurs in a similar way for several consecutive peaks
\footnote{At relatively high temperatures the conductance of a few consecutive peaks in  \cite{Meirav} does decrease as the temperature is decreased, but the peak to valley difference increases, as opposed to our results, cf. Fig.~S3 of the supplementary material \cite{SMArchive}}.

Now we address the question of why in our InAs NWs the effective LL parameter $g$ is smaller than $1/2$ at low filling (so that $G_\text{max}$ \emph{decreases} with decreasing temperature), while other experimental studies of 1D quantum wires, \textit{e.g.}, carbon nanotubes \cite{nanotubeLL}, GaAs wires formed at the cleaved edge overgrowth of a GaAs/AlGaAs heterostructure \cite{Yacobi}, or at the bottom of a V-grooved GaAs growth \cite{Vgroove}, all exhibit effective LL parameters higher than $1/2$ (so that $G_\text{max}$ \emph{increases} with decreasing temperature).
We believe that there are two main reasons which contribute to the lower value of the LL parameter in our InAs NWs.

The first is related to the environment of the quantum wires. Both types of GaAs wires reported in the literature \cite{Yacobi,Vgroove} were created within 2DEG structures embedded well inside a semiconductor material with a large dielectric constant (AlGaAs and GaAs). In contrast, while the InAs NWs have a similar dielectric constant to GaAs, they are placed on \SiO~surface, so the surrounding materials, namely, $^4$He and \SiO, possess much smaller dielectric constants. These reduced dielectric constants enhance the effect of Coulomb interaction in our system as compared to the GaAs wires  reported before.. 
Carbon nanotubes could have even lower dielectric constant (especially when suspended), but, as we now discuss, their high channel symmetry is responsible for enhancing the effective $g$.

The second reason for observing a smaller LL parameter in our system is related to an inherent property of InAs, which possesses strong spin-orbit coupling that breaks the spin rotation symmetry. It should be noted that the effective LL parameter $g$ is related to the interaction parameters in the charge and spin channels ($g_c$ and $g_s$, respectively) by \cite{paper10}
\begin{equation}
\frac{1}{g}=\frac{1}{2g_{c}}+\frac{1}{2g_{s}}.
\label{eq5}
\end{equation}
In GaAs spin-orbit coupling is very small, therefore spin-rotation symmetry dictates that $g_s=1$. Thus, $g<1/2$ can only be obtained if the interaction in the charge sector is extremely strong, $g_c<1/4$. In carbon nanotubes the additional valley degeneracy results in $1/g = 1/(4g_c)+3/4$, so $g<1/2$ results in an even stricter condition, $g_c<1/5$.
On the other hand, in InAs spin rotation symmetry is broken, allowing for $g_s<1$ and making it easier to reach $g<1/2$.

Therefore  we believe that the combination of a lack of orbital degeneracy due to strong spin-orbit coupling and of the low effective dielectric constant, makes our InAs NW a unique system where strong effective interactions, $g<1/2$, can be achieved, and thus a decrease in Coulomb blockade peak heights with decreasing temperature can be observed.
Finally, our experiment concentrated on the sequential tunneling regime, at lower temperature we expect resonant tunneling to become dominant near the Coulomb blockade peaks, and the Kondo effect to show up in odd charge valleys, as recently found in measurements carried out on short InAs NWs (where LL effects are absent) \cite{Kretinin}.

\begin{acknowledgments}
We are thankful to Ronit Popotitz-Biro for professional TEM study of the InAs NWs.
We gratefully acknowledge support by the ISF BIKURA program and GIF (MG); ISF and Marie Curie CIG grants (ES); as well as ISF grant \#532/12 and IMOST grants \#3-11173 (AP and HS) \& \#3-8668 (HS).
\end{acknowledgments}

\bibliography{InAsLLarticlebib}

\begin{thebibliography}{21}%
\makeatletter
\providecommand \@ifxundefined [1]{%
 \@ifx{#1\undefined}
}%
\providecommand \@ifnum [1]{%
 \ifnum #1\expandafter \@firstoftwo
 \else \expandafter \@secondoftwo
 \fi
}%
\providecommand \@ifx [1]{%
 \ifx #1\expandafter \@firstoftwo
 \else \expandafter \@secondoftwo
 \fi
}%
\providecommand \natexlab [1]{#1}%
\providecommand \enquote  [1]{``#1''}%
\providecommand \bibnamefont  [1]{#1}%
\providecommand \bibfnamefont [1]{#1}%
\providecommand \citenamefont [1]{#1}%
\providecommand \href@noop [0]{\@secondoftwo}%
\providecommand \href [0]{\begingroup \@sanitize@url \@href}%
\providecommand \@href[1]{\@@startlink{#1}\@@href}%
\providecommand \@@href[1]{\endgroup#1\@@endlink}%
\providecommand \@sanitize@url [0]{\catcode `\\12\catcode `\$12\catcode
  `\&12\catcode `\#12\catcode `\^12\catcode `\_12\catcode `\%12\relax}%
\providecommand \@@startlink[1]{}%
\providecommand \@@endlink[0]{}%
\providecommand \url  [0]{\begingroup\@sanitize@url \@url }%
\providecommand \@url [1]{\endgroup\@href {#1}{\urlprefix }}%
\providecommand \urlprefix  [0]{URL }%
\providecommand \Eprint [0]{\href }%
\providecommand \doibase [0]{http://dx.doi.org/}%
\providecommand \selectlanguage [0]{\@gobble}%
\providecommand \bibinfo  [0]{\@secondoftwo}%
\providecommand \bibfield  [0]{\@secondoftwo}%
\providecommand \translation [1]{[#1]}%
\providecommand \BibitemOpen [0]{}%
\providecommand \bibitemStop [0]{}%
\providecommand \bibitemNoStop [0]{.\EOS\space}%
\providecommand \EOS [0]{\spacefactor3000\relax}%
\providecommand \BibitemShut  [1]{\csname bibitem#1\endcsname}%
\let\auto@bib@innerbib\@empty
\bibitem [{\citenamefont {Lutchyn}\ \emph {et~al.}(2010)\citenamefont
  {Lutchyn}, \citenamefont {Sau},\ and\ \citenamefont {Das~Sarma}}]{paper1}%
  \BibitemOpen
  \bibfield  {author} {\bibinfo {author} {\bibfnamefont {R.~M.}\ \bibnamefont
  {Lutchyn}}, \bibinfo {author} {\bibfnamefont {J.~D.}\ \bibnamefont {Sau}}, \
  and\ \bibinfo {author} {\bibfnamefont {S.}~\bibnamefont {Das~Sarma}},\ }\href
  {\doibase 10.1103/PhysRevLett.105.077001} {\bibfield  {journal} {\bibinfo
  {journal} {Phys. Rev. Lett.}\ }\textbf {\bibinfo {volume} {105}},\ \bibinfo
  {pages} {077001} (\bibinfo {year} {2010})}\BibitemShut {NoStop}%
\bibitem [{\citenamefont {Oreg}\ \emph {et~al.}(2010)\citenamefont {Oreg},
  \citenamefont {Refael},\ and\ \citenamefont {von Oppen}}]{paper2}%
  \BibitemOpen
  \bibfield  {author} {\bibinfo {author} {\bibfnamefont {Y.}~\bibnamefont
  {Oreg}}, \bibinfo {author} {\bibfnamefont {G.}~\bibnamefont {Refael}}, \ and\
  \bibinfo {author} {\bibfnamefont {F.}~\bibnamefont {von Oppen}},\ }\href
  {\doibase 10.1103/PhysRevLett.105.177002} {\bibfield  {journal} {\bibinfo
  {journal} {Phys. Rev. Lett.}\ }\textbf {\bibinfo {volume} {105}},\ \bibinfo
  {pages} {177002} (\bibinfo {year} {2010})}\BibitemShut {NoStop}%
\bibitem [{\citenamefont {Das}\ \emph {et~al.}(2012)\citenamefont {Das},
  \citenamefont {Ronen}, \citenamefont {Most}, \citenamefont {Oreg},
  \citenamefont {Heiblum},\ and\ \citenamefont {Shtrikman}}]{paper3}%
  \BibitemOpen
  \bibfield  {author} {\bibinfo {author} {\bibfnamefont {A.}~\bibnamefont
  {Das}}, \bibinfo {author} {\bibfnamefont {Y.}~\bibnamefont {Ronen}}, \bibinfo
  {author} {\bibfnamefont {Y.}~\bibnamefont {Most}}, \bibinfo {author}
  {\bibfnamefont {Y.}~\bibnamefont {Oreg}}, \bibinfo {author} {\bibfnamefont
  {M.}~\bibnamefont {Heiblum}}, \ and\ \bibinfo {author} {\bibfnamefont
  {H.}~\bibnamefont {Shtrikman}},\ }\href@noop {} {\bibfield  {journal}
  {\bibinfo  {journal} {Nat. Phys.}\ }\textbf {\bibinfo {volume} {8}},\
  \bibinfo {pages} {887} (\bibinfo {year} {2012})}\BibitemShut {NoStop}%
\bibitem [{\citenamefont {Mourik}\ \emph {et~al.}(2012)\citenamefont {Mourik},
  \citenamefont {Zuo}, \citenamefont {Frolov}, \citenamefont {Plissard},
  \citenamefont {Bakkers},\ and\ \citenamefont {Kouwenhoven}}]{paper4}%
  \BibitemOpen
  \bibfield  {author} {\bibinfo {author} {\bibfnamefont {V.}~\bibnamefont
  {Mourik}}, \bibinfo {author} {\bibfnamefont {K.}~\bibnamefont {Zuo}},
  \bibinfo {author} {\bibfnamefont {S.}~\bibnamefont {Frolov}}, \bibinfo
  {author} {\bibfnamefont {S.}~\bibnamefont {Plissard}}, \bibinfo {author}
  {\bibfnamefont {E.}~\bibnamefont {Bakkers}}, \ and\ \bibinfo {author}
  {\bibfnamefont {L.}~\bibnamefont {Kouwenhoven}},\ }\href@noop {} {\bibfield
  {journal} {\bibinfo  {journal} {Science}\ }\textbf {\bibinfo {volume}
  {336}},\ \bibinfo {pages} {1003} (\bibinfo {year} {2012})}\BibitemShut
  {NoStop}%
\bibitem [{\citenamefont {Deng}\ \emph {et~al.}(2012)\citenamefont {Deng},
  \citenamefont {Yu}, \citenamefont {Huang}, \citenamefont {Larsson},
  \citenamefont {Caroff},\ and\ \citenamefont {Xu}}]{paper5}%
  \BibitemOpen
  \bibfield  {author} {\bibinfo {author} {\bibfnamefont {M.}~\bibnamefont
  {Deng}}, \bibinfo {author} {\bibfnamefont {C.}~\bibnamefont {Yu}}, \bibinfo
  {author} {\bibfnamefont {G.}~\bibnamefont {Huang}}, \bibinfo {author}
  {\bibfnamefont {M.}~\bibnamefont {Larsson}}, \bibinfo {author} {\bibfnamefont
  {P.}~\bibnamefont {Caroff}}, \ and\ \bibinfo {author} {\bibfnamefont
  {H.}~\bibnamefont {Xu}},\ }\href@noop {} {\bibfield  {journal} {\bibinfo
  {journal} {Nano Lett.}\ }\textbf {\bibinfo {volume} {12}},\ \bibinfo {pages}
  {6414} (\bibinfo {year} {2012})}\BibitemShut {NoStop}%
\bibitem [{\citenamefont {Pershin}\ \emph {et~al.}(2004)\citenamefont
  {Pershin}, \citenamefont {Nesteroff},\ and\ \citenamefont
  {Privman}}]{paper6}%
  \BibitemOpen
  \bibfield  {author} {\bibinfo {author} {\bibfnamefont {Y.~V.}\ \bibnamefont
  {Pershin}}, \bibinfo {author} {\bibfnamefont {J.~A.}\ \bibnamefont
  {Nesteroff}}, \ and\ \bibinfo {author} {\bibfnamefont {V.}~\bibnamefont
  {Privman}},\ }\href {\doibase 10.1103/PhysRevB.69.121306} {\bibfield
  {journal} {\bibinfo  {journal} {Phys. Rev. B}\ }\textbf {\bibinfo {volume}
  {69}},\ \bibinfo {pages} {121306} (\bibinfo {year} {2004})}\BibitemShut
  {NoStop}%
\bibitem [{\citenamefont {Krive}\ \emph {et~al.}(2010)\citenamefont {Krive},
  \citenamefont {Palevski}, \citenamefont {Shekhter},\ and\ \citenamefont
  {Jonson}}]{paper8}%
  \BibitemOpen
  \bibfield  {author} {\bibinfo {author} {\bibfnamefont {I.}~\bibnamefont
  {Krive}}, \bibinfo {author} {\bibfnamefont {A.}~\bibnamefont {Palevski}},
  \bibinfo {author} {\bibfnamefont {R.}~\bibnamefont {Shekhter}}, \ and\
  \bibinfo {author} {\bibfnamefont {M.}~\bibnamefont {Jonson}},\ }\href@noop {}
  {\bibfield  {journal} {\bibinfo  {journal} {Fiz. Nizk. Temp.}\ }\textbf
  {\bibinfo {volume} {36}},\ \bibinfo {pages} {155} (\bibinfo {year}
  {2010})}\BibitemShut {NoStop}%
\bibitem [{\citenamefont {Field}\ \emph {et~al.}(1990)\citenamefont {Field},
  \citenamefont {Kastner}, \citenamefont {Meirav}, \citenamefont
  {Scott-Thomas}, \citenamefont {Antoniadis}, \citenamefont {Smith},\ and\
  \citenamefont {Wind}}]{Meirav}%
  \BibitemOpen
  \bibfield  {author} {\bibinfo {author} {\bibfnamefont {S.~B.}\ \bibnamefont
  {Field}}, \bibinfo {author} {\bibfnamefont {M.~A.}\ \bibnamefont {Kastner}},
  \bibinfo {author} {\bibfnamefont {U.}~\bibnamefont {Meirav}}, \bibinfo
  {author} {\bibfnamefont {J.~H.~F.}\ \bibnamefont {Scott-Thomas}}, \bibinfo
  {author} {\bibfnamefont {D.~A.}\ \bibnamefont {Antoniadis}}, \bibinfo
  {author} {\bibfnamefont {H.~I.}\ \bibnamefont {Smith}}, \ and\ \bibinfo
  {author} {\bibfnamefont {S.~J.}\ \bibnamefont {Wind}},\ }\href {\doibase
  10.1103/PhysRevB.42.3523} {\bibfield  {journal} {\bibinfo  {journal} {Phys.
  Rev. B}\ }\textbf {\bibinfo {volume} {42}},\ \bibinfo {pages} {3523}
  (\bibinfo {year} {1990})}\BibitemShut {NoStop}%
\bibitem [{\citenamefont {Moser}\ \emph {et~al.}(2006)\citenamefont {Moser},
  \citenamefont {Roddaro}, \citenamefont {Schuh}, \citenamefont {Bichler},
  \citenamefont {Pellegrini},\ and\ \citenamefont {Grayson}}]{Moser}%
  \BibitemOpen
  \bibfield  {author} {\bibinfo {author} {\bibfnamefont {J.}~\bibnamefont
  {Moser}}, \bibinfo {author} {\bibfnamefont {S.}~\bibnamefont {Roddaro}},
  \bibinfo {author} {\bibfnamefont {D.}~\bibnamefont {Schuh}}, \bibinfo
  {author} {\bibfnamefont {M.}~\bibnamefont {Bichler}}, \bibinfo {author}
  {\bibfnamefont {V.}~\bibnamefont {Pellegrini}}, \ and\ \bibinfo {author}
  {\bibfnamefont {M.}~\bibnamefont {Grayson}},\ }\href {\doibase
  10.1103/PhysRevB.74.193307} {\bibfield  {journal} {\bibinfo  {journal} {Phys.
  Rev. B}\ }\textbf {\bibinfo {volume} {74}},\ \bibinfo {pages} {193307}
  (\bibinfo {year} {2006})}\BibitemShut {NoStop}%
\bibitem [{\citenamefont {Ruzin}\ \emph {et~al.}(1992)\citenamefont {Ruzin},
  \citenamefont {Chandrasekhar}, \citenamefont {Levin},\ and\ \citenamefont
  {Glazman}}]{Glazman}%
  \BibitemOpen
  \bibfield  {author} {\bibinfo {author} {\bibfnamefont {I.~M.}\ \bibnamefont
  {Ruzin}}, \bibinfo {author} {\bibfnamefont {V.}~\bibnamefont
  {Chandrasekhar}}, \bibinfo {author} {\bibfnamefont {E.~I.}\ \bibnamefont
  {Levin}}, \ and\ \bibinfo {author} {\bibfnamefont {L.~I.}\ \bibnamefont
  {Glazman}},\ }\href {\doibase 10.1103/PhysRevB.45.13469} {\bibfield
  {journal} {\bibinfo  {journal} {Phys. Rev. B}\ }\textbf {\bibinfo {volume}
  {45}},\ \bibinfo {pages} {13469} (\bibinfo {year} {1992})}\BibitemShut
  {NoStop}%
\bibitem [{\citenamefont {Shtrikman}\ \emph {et~al.}(2011)\citenamefont
  {Shtrikman}, \citenamefont {Popovitz-Biro}, \citenamefont {Kretinin},\ and\
  \citenamefont {Kacman}}]{paper9}%
  \BibitemOpen
  \bibfield  {author} {\bibinfo {author} {\bibfnamefont {H.}~\bibnamefont
  {Shtrikman}}, \bibinfo {author} {\bibfnamefont {R.}~\bibnamefont
  {Popovitz-Biro}}, \bibinfo {author} {\bibfnamefont {A.}~\bibnamefont
  {Kretinin}}, \ and\ \bibinfo {author} {\bibfnamefont {P.}~\bibnamefont
  {Kacman}},\ }\href@noop {} {\bibfield  {journal} {\bibinfo  {journal} {IEEE
  J. Selected Top. Quant. Electron}\ }\textbf {\bibinfo {volume} {17}},\
  \bibinfo {pages} {992} (\bibinfo {year} {2011})}\BibitemShut {NoStop}%
\bibitem [{\citenamefont {Suyatin}\ \emph {et~al.}(2007)\citenamefont
  {Suyatin}, \citenamefont {Thelander}, \citenamefont {Bj\"ork}, \citenamefont
  {Maximov},\ and\ \citenamefont {Samuelson}}]{Ammoniumsolfide}%
  \BibitemOpen
  \bibfield  {author} {\bibinfo {author} {\bibfnamefont {D.~B.}\ \bibnamefont
  {Suyatin}}, \bibinfo {author} {\bibfnamefont {C.}~\bibnamefont {Thelander}},
  \bibinfo {author} {\bibfnamefont {M.~T.}\ \bibnamefont {Bj\"ork}}, \bibinfo
  {author} {\bibfnamefont {I.}~\bibnamefont {Maximov}}, \ and\ \bibinfo
  {author} {\bibfnamefont {L.}~\bibnamefont {Samuelson}},\ }\href@noop {}
  {\bibfield  {journal} {\bibinfo  {journal} {Nanotechnology}\ }\textbf
  {\bibinfo {volume} {18}},\ \bibinfo {pages} {105307} (\bibinfo {year}
  {2007})}\BibitemShut {NoStop}%
\bibitem [{\citenamefont {Pustilnik}\ and\ \citenamefont
  {Glazman}(2004)}]{QDreview}%
  \BibitemOpen
  \bibfield  {author} {\bibinfo {author} {\bibfnamefont {M.}~\bibnamefont
  {Pustilnik}}\ and\ \bibinfo {author} {\bibfnamefont {L.}~\bibnamefont
  {Glazman}},\ }\href@noop {} {\bibfield  {journal} {\bibinfo  {journal} {J.
  Phys.: Condens. Matter}\ }\textbf {\bibinfo {volume} {16}},\ \bibinfo {pages}
  {R513} (\bibinfo {year} {2004})}\BibitemShut {NoStop}%
\bibitem [{\citenamefont {Furusaki}(1998)}]{paper10}%
  \BibitemOpen
  \bibfield  {author} {\bibinfo {author} {\bibfnamefont {A.}~\bibnamefont
  {Furusaki}},\ }\href {\doibase 10.1103/PhysRevB.57.7141} {\bibfield
  {journal} {\bibinfo  {journal} {Phys. Rev. B}\ }\textbf {\bibinfo {volume}
  {57}},\ \bibinfo {pages} {7141} (\bibinfo {year} {1998})}\BibitemShut
  {NoStop}%
\bibitem [{SMA()}]{SMArchive}%
  \BibitemOpen
  \href@noop {} {\bibinfo  {journal} {See supplementary material on page 6 for
  additional experimental data}\ }\BibitemShut {NoStop}%
\bibitem [{\citenamefont {Meng}\ \emph {et~al.}(2014)\citenamefont {Meng},
  \citenamefont {Klinovaja},\ and\ \citenamefont {Loss}}]{Meng}%
  \BibitemOpen
\bibfield  {journal} {  }\bibfield  {author} {\bibinfo {author} {\bibfnamefont
  {T.}~\bibnamefont {Meng}}, \bibinfo {author} {\bibfnamefont {J.}~\bibnamefont
  {Klinovaja}}, \ and\ \bibinfo {author} {\bibfnamefont {D.}~\bibnamefont
  {Loss}},\ }\href {\doibase 10.1103/PhysRevB.89.205133} {\bibfield  {journal}
  {\bibinfo  {journal} {Phys. Rev. B}\ }\textbf {\bibinfo {volume} {89}},\
  \bibinfo {pages} {205133} (\bibinfo {year} {2014})}\BibitemShut {NoStop}%
\bibitem [{Note1()}]{Note1}%
  \BibitemOpen
  \bibinfo {note} {At relatively high temperatures the conductance of a few
  consecutive peaks in \cite {Meirav} does decrease as the temperature is
  decreased, but the peak to valley difference increases, as opposed to our
  results, cf. Fig.~S3 of the supplementary material \cite
  {SMArchive}}\BibitemShut {NoStop}%
\bibitem [{\citenamefont {Bockrath}\ \emph {et~al.}(1999)\citenamefont
  {Bockrath}, \citenamefont {Cobden}, \citenamefont {Lu}, \citenamefont
  {Rinzler}, \citenamefont {Smalley}, \citenamefont {Balents},\ and\
  \citenamefont {McEuen}}]{nanotubeLL}%
  \BibitemOpen
  \bibfield  {author} {\bibinfo {author} {\bibfnamefont {M.}~\bibnamefont
  {Bockrath}}, \bibinfo {author} {\bibfnamefont {D.~H.}\ \bibnamefont
  {Cobden}}, \bibinfo {author} {\bibfnamefont {J.}~\bibnamefont {Lu}}, \bibinfo
  {author} {\bibfnamefont {A.~G.}\ \bibnamefont {Rinzler}}, \bibinfo {author}
  {\bibfnamefont {R.~E.}\ \bibnamefont {Smalley}}, \bibinfo {author}
  {\bibfnamefont {L.}~\bibnamefont {Balents}}, \ and\ \bibinfo {author}
  {\bibfnamefont {P.~L.}\ \bibnamefont {McEuen}},\ }\href@noop {} {\bibfield
  {journal} {\bibinfo  {journal} {Nature}\ }\textbf {\bibinfo {volume} {397}},\
  \bibinfo {pages} {598} (\bibinfo {year} {1999})}\BibitemShut {NoStop}%
\bibitem [{\citenamefont {Auslaender}\ \emph {et~al.}(2000)\citenamefont
  {Auslaender}, \citenamefont {Yacoby}, \citenamefont {de~Picciotto},
  \citenamefont {Baldwin}, \citenamefont {Pfeiffer},\ and\ \citenamefont
  {West}}]{Yacobi}%
  \BibitemOpen
  \bibfield  {author} {\bibinfo {author} {\bibfnamefont {O.~M.}\ \bibnamefont
  {Auslaender}}, \bibinfo {author} {\bibfnamefont {A.}~\bibnamefont {Yacoby}},
  \bibinfo {author} {\bibfnamefont {R.}~\bibnamefont {de~Picciotto}}, \bibinfo
  {author} {\bibfnamefont {K.~W.}\ \bibnamefont {Baldwin}}, \bibinfo {author}
  {\bibfnamefont {L.~N.}\ \bibnamefont {Pfeiffer}}, \ and\ \bibinfo {author}
  {\bibfnamefont {K.~W.}\ \bibnamefont {West}},\ }\href {\doibase
  10.1103/PhysRevLett.84.1764} {\bibfield  {journal} {\bibinfo  {journal}
  {Phys. Rev. Lett.}\ }\textbf {\bibinfo {volume} {84}},\ \bibinfo {pages}
  {1764} (\bibinfo {year} {2000})}\BibitemShut {NoStop}%
\bibitem [{\citenamefont {Levy}\ \emph {et~al.}(2012)\citenamefont {Levy},
  \citenamefont {Sternfeld}, \citenamefont {Eshkol}, \citenamefont {Karpovski},
  \citenamefont {Dwir}, \citenamefont {Rudra}, \citenamefont {Kapon},
  \citenamefont {Oreg},\ and\ \citenamefont {Palevski}}]{Vgroove}%
  \BibitemOpen
  \bibfield  {author} {\bibinfo {author} {\bibfnamefont {E.}~\bibnamefont
  {Levy}}, \bibinfo {author} {\bibfnamefont {I.}~\bibnamefont {Sternfeld}},
  \bibinfo {author} {\bibfnamefont {M.}~\bibnamefont {Eshkol}}, \bibinfo
  {author} {\bibfnamefont {M.}~\bibnamefont {Karpovski}}, \bibinfo {author}
  {\bibfnamefont {B.}~\bibnamefont {Dwir}}, \bibinfo {author} {\bibfnamefont
  {A.}~\bibnamefont {Rudra}}, \bibinfo {author} {\bibfnamefont
  {E.}~\bibnamefont {Kapon}}, \bibinfo {author} {\bibfnamefont
  {Y.}~\bibnamefont {Oreg}}, \ and\ \bibinfo {author} {\bibfnamefont
  {A.}~\bibnamefont {Palevski}},\ }\href {\doibase 10.1103/PhysRevB.85.045315}
  {\bibfield  {journal} {\bibinfo  {journal} {Phys. Rev. B}\ }\textbf {\bibinfo
  {volume} {85}},\ \bibinfo {pages} {045315} (\bibinfo {year}
  {2012})}\BibitemShut {NoStop}%
\bibitem [{\citenamefont {Kretinin}\ \emph {et~al.}(2011)\citenamefont
  {Kretinin}, \citenamefont {Shtrikman}, \citenamefont {Goldhaber-Gordon},
  \citenamefont {Hanl}, \citenamefont {Weichselbaum}, \citenamefont {von
  Delft}, \citenamefont {Costi},\ and\ \citenamefont {Mahalu}}]{Kretinin}%
  \BibitemOpen
  \bibfield  {author} {\bibinfo {author} {\bibfnamefont {A.~V.}\ \bibnamefont
  {Kretinin}}, \bibinfo {author} {\bibfnamefont {H.}~\bibnamefont {Shtrikman}},
  \bibinfo {author} {\bibfnamefont {D.}~\bibnamefont {Goldhaber-Gordon}},
  \bibinfo {author} {\bibfnamefont {M.}~\bibnamefont {Hanl}}, \bibinfo {author}
  {\bibfnamefont {A.}~\bibnamefont {Weichselbaum}}, \bibinfo {author}
  {\bibfnamefont {J.}~\bibnamefont {von Delft}}, \bibinfo {author}
  {\bibfnamefont {T.}~\bibnamefont {Costi}}, \ and\ \bibinfo {author}
  {\bibfnamefont {D.}~\bibnamefont {Mahalu}},\ }\href {\doibase
  10.1103/PhysRevB.84.245316} {\bibfield  {journal} {\bibinfo  {journal} {Phys.
  Rev. B}\ }\textbf {\bibinfo {volume} {84}},\ \bibinfo {pages} {245316}
  (\bibinfo {year} {2011})}\BibitemShut {NoStop}%
\end{thebibliography}%
\newpage
\maketitle

\beginsupplement
\begin{figure}{}
\centering
\includegraphics[width=0.38\textwidth]{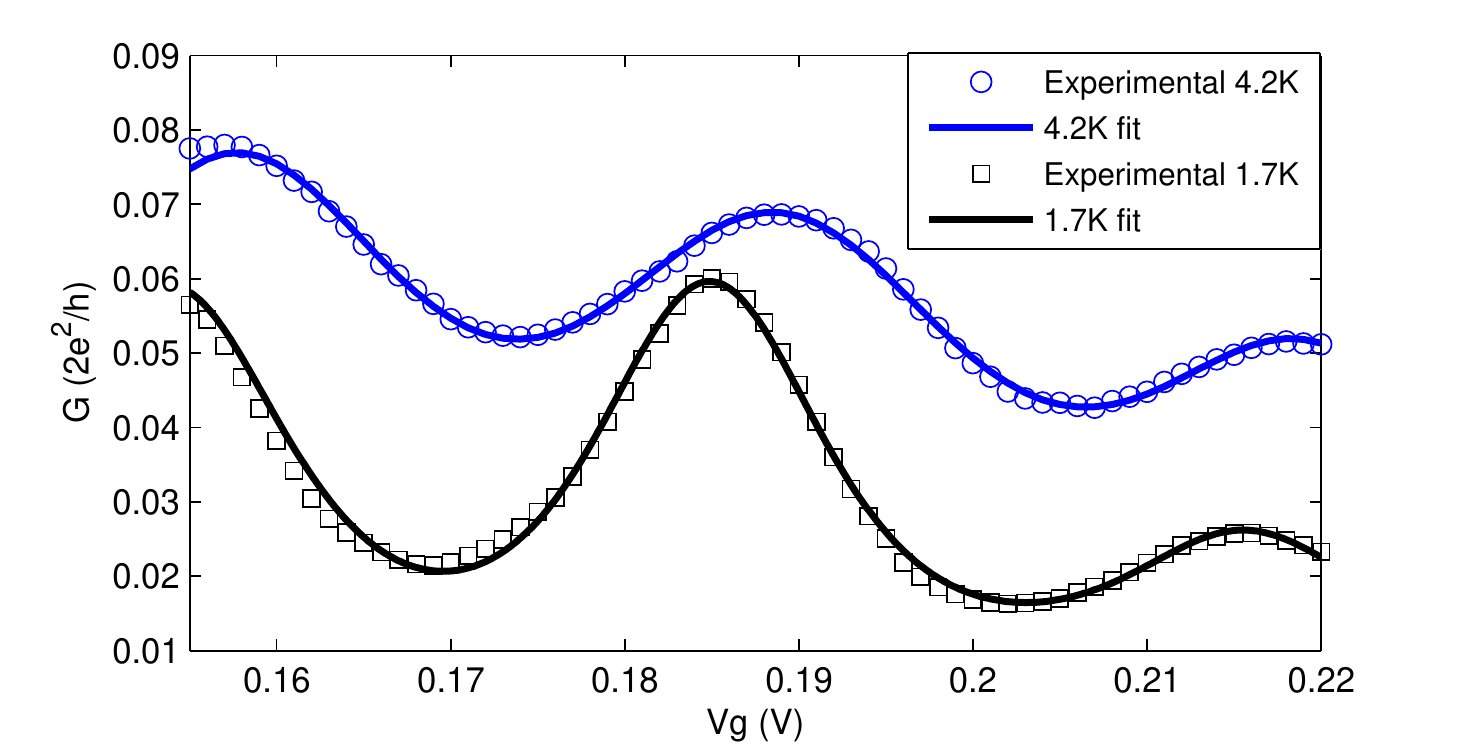}
\caption{Fit of peaks \#16--\#18 in Fig.~2 to Eq.~(1) with $\alpha=0.08$ and $g=0.5$ at 4.2~K and 1.7~K. }\label{SM1}
\centering
\includegraphics[width=0.38\textwidth]{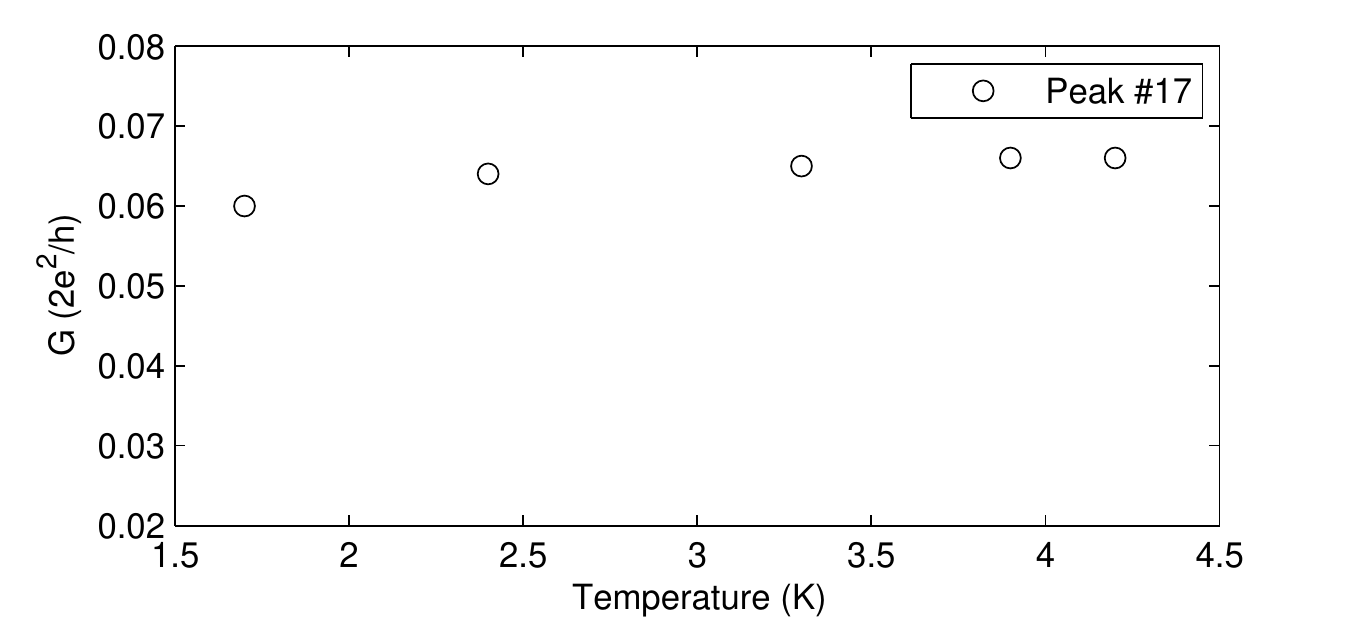}
\caption{The maximum conductance of peak \#17 versus temperature, as was extracted from the fitting to Eq.~(1) in Fig.~S1. The peak height is almost constant as function of temperature, indicating $g \approx 0.5$ [cf. Eq. (2)].}\label{SM2}
\end{figure}
\begin{figure}{}
\centering
\includegraphics[width=0.38\textwidth]{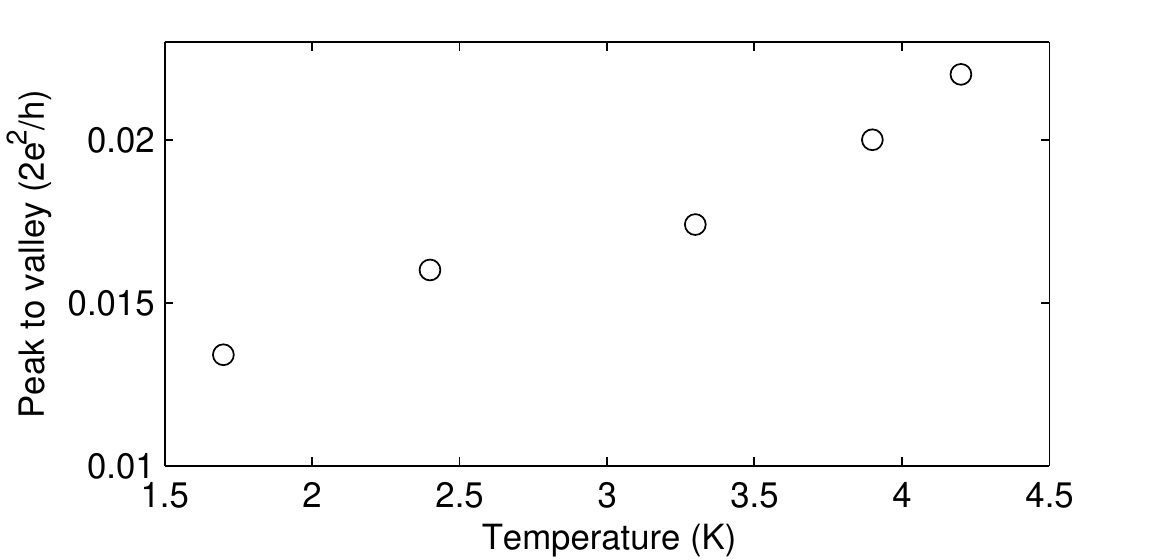}
\caption{The difference between the peak and the subsequent valley conductance as function of temperature for the first peak in Fig.~2. As discussed in footnote 17 in the main text this behavior is opposite to the one observed in Ref. [8] where the peak to valley distance decreases with temperature.}\label{SM3}
\end{figure}

\end{document}